\documentclass[
 aps,prb,twocolumn,showpacs,
twoside,superscriptaddress,nofootinbib,amsmath,amssymb
]{revtex4-2}

\usepackage{graphicx}
\usepackage{dcolumn}
\usepackage{bm}
\usepackage{hyperref}
\usepackage{physics}
\usepackage{makecell}
\usepackage{xcolor}
\usepackage{tabularx}
\usepackage[caption=false, font=large]{subfig}
\usepackage{tikz}
\usepackage{circuitikz}
\usepackage[mathlines]{lineno}
 
\begin{document}

\preprint{APS/123-QED}

\title{Quantum dynamics of frustrated Josephson junction arrays embedded in a transmission line: an effective $XX$ spin chain with long-range interaction}

\author{Benedikt J.P. Pernack}
 \email{Benedikt.Pernack@ruhr-uni-bochum.de}
\author{Mikhail V. Fistul}
\author{Ilya M. Eremin}  
\affiliation{
 Theoretische Physik III, Ruhr-Universit\"at Bochum, Bochum 44801, Germany
}

\date{\today}

\begin{abstract}
We study theoretically a variety of collective quantum phases occurring in frustrated saw-tooth chains of Josephson junctions embedded in a dissipationless transmission line. The basic element of a system, i.e., the triangular superconducting cell, contains two $0$- and one $\pi$- Josephson junctions characterized by  $E_J$ and $\alpha E_J$ Josephson energies, accordingly. In the frustrated regime the low energy quantum dynamics of a single cell is determined by anticlockwise or clockwise flowing persistent currents (vortex/antivortex). The direct embedding of $\pi$-Josephson junctions in a transmission line allows to establish a short/long-range interaction between (anti)vortices of well separated cells. By making use of the variational approach, we map the superconducting circuit Hamiltonian to an effective $XX$ spin model with an exchange spin-spin interaction decaying with the distance $x$ as $x^{-\beta}$, and the local $\hat \sigma_{x,n}$-terms corresponding to the coherent quantum beats between vortex and antivortex in a single cell. We obtain that in long arrays as $N \gg \ell_0 \simeq \sqrt{C/C_0}$, where $C$ and $C_0$ are capacitances of $0$-Josephson junction and transmission line, accordingly, the amplitude of quantum beats is strongly suppressed. By means of exact numerical diagonalization, we study the interplay between the coherent quantum beats and the exchange spin-spin interaction leading to the appearance of various collective quantum phases such as the paramagnetic ($P$), compressible superfluid ($CS$) and weakly compressible superfluid ($w$-$CS$) states.
\end{abstract}

\maketitle

\section{\label{sec1:level1}Introduction}

Different kinds of \textit{frustrated} systems and their novel physics have motivated many experiments and theoretical studies in recent times. Initially, the concept of frustration was introduced in magnetic materials, in which either geometric frustration or frustration emerging due to competing interactions, e.g., ferromagnetic/anti-ferromagnetic interactions, occur  \cite{Anderson1978, Moessner2006, Balents2010, Baniodeh2018}. The unique properties of frustrated systems are a highly degenerated ground state, multiple low-lying metastable states with long relaxation times at low temperatures \cite{Moessner2006, Baniodeh2018, Richter2018}. 

In order to give a short flavour of the richness of the collective phases and phase transitions realized in frustrated systems, we refer to the discovery of nematic order in ferromagnetic superconductors \cite{Gong2017, Fernandes2014} and frustrated ferromagnetic chains \cite{Zhitomirsky2010}, non-collinear ferrimagnet behaviour \cite{Baniodeh2018},  magnetic order-disorder phase transitions \cite{Richter2018}, spin-liquids \cite{Moessner2006, Tian-Heng2012, Teitel1983, Balents2010} and topological vortices \cite{Moessner2006, Kitaev2003}. Such novel phases and phase transitions are observed in strongly correlated electronic systems \cite{Moessner2006, Zhitomirsky2010, Richter2018, Caretta2006}, e.g., in kagome superconductors \cite{Wang2021}, in antiferromagnets \cite{Tian-Heng2012, Yamada2016, Messio2012, Chernyshev2014}, in quantum magnets \cite{Teitel1983, Kitaev2003} and also in quasi-one-dimensional molecular systems displaying a transition to a chiral spin-liquid state \cite{Cinti2008}. 

The frustration has been implemented in various artificial intrinsically quantum platforms such as trapped ion systems \cite{Britton2012, Monroe2021, Lewis2023}, photonic crystals \cite{Weimann2016, Vincencio2015}, Rydberg arrays \cite{Semeghini2021, Giudici2022, Chen2022} and Josephson junction arrays (JJAs) \cite{Pop2008, Johnson2011, King2018, King2021}. In the latter systems, produced in very different geometries and sizes, fascinating physical effects, e.g., nonlinear classical dynamics of magnetic Josephson vortices (fluxons) \cite{ustinov1993fluxon,wallraff2003quantum}, discrete breathers in Josephson junction ladders \cite{trias2000discrete,binder2000observation}, superconductor-insulator quantum phase transitions \cite{Haviland2000, Doucot2002} have been intensively studied experimentally and theoretically in last decades.  
Furthermore, Josephson junction arrays provide a promising platform for future realizations of different superconducting Josephson meta-materials in the context of novel quantum devices and technology \cite{Jung2014, Shulga2018, Ranadive2022}. These quantum devices might be suitable for realization of analog quantum simulations in different fields of quantum science \cite{georgescu2014quantum,Buluta2009,acin2018quantum,Daley2022}.

There is a particular interest in frustrated Josephson junction arrays ($f$-JJAs) with at least two different ways how the frustration can be introduced in \textit{f}-JJAs: (i) application of an external magnetic field \cite{Orlando1999, Doucot2002, Pop2008} and (ii) usage of a combination of $0$ and $\pi$-Josephson junctions implemented in a single triangular unit cell \cite{Fistul2019, Neyenhuys2023}. Depending on the frustration parameter, 
\textit{f}-JJAs exhibits either frustrated or non-frustrated regime. In the frustrated regime, for a single triangular unit cell, the potential energy shows a double well shape with two degenerate minima corresponding to the anticlockwise or clockwise flowing persistent currents (vortex/antivortex) \cite{Orlando1999}. 
This generalizes to a more complex square and triangular lattices of \textit{f}-JJAs, where highly degenerated complex ground states corresponding to the different configurations of vortices/antivortices, e.g., checkerboard, ribbon or stripe types have been observed \cite{Caputo2001, benz1990critical, Valdez2005}.
\begin{figure*}
    \centering	
	\includegraphics[width=0.83\textwidth]{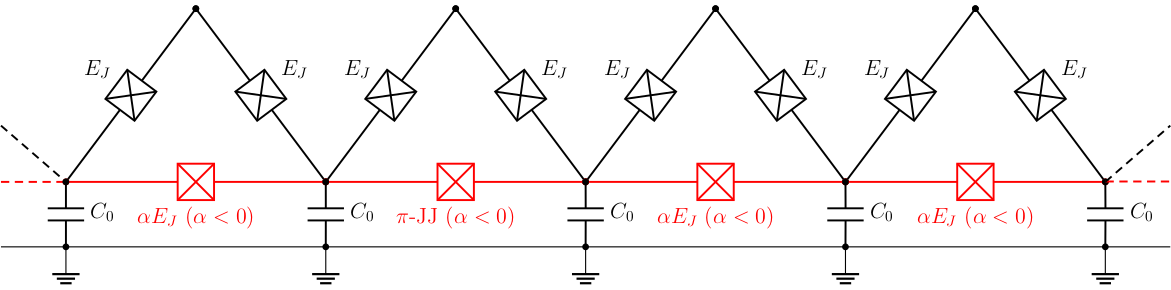}   
    \caption{\label{fig:one} (color online) Schematic of a frustrated saw-tooth chain of Josephson junctions directly embedded in a dissipationless transmission line characterized by the capacitance to the ground $C_0$. The $0$- and $\pi$- Josephson junctions are shown in black (red). $E_J$ and $\alpha E_J$ are the Josephson energies of $0$- and $\pi$ Josephson junctions, respectively. 
}
\end{figure*}
The coherent \textit{quantum} dynamics of a single triangular unit cell biased in the frustrated regime is determined by macroscopic quantum tunneling between two degenerate minima leading to the coherent quantum beats between the anti-clockwise/clockwise flowing persistent currents. Such a lump superconducting quantum circuit is identical to a single flux qubit biased to the symmetry point \cite{Orlando1999}.

Single triangular unit cells can be arranged in plenty different \textit{vertex-sharing} $f$-JJAs: quasi-one-dimensional systems such as saw-tooth and diamond chains \cite{Pop2008, Doucot2002, Rizzi2006, Fistul2019} as well as a two dimensional Kagome lattice \cite{Fistul2020, Neyenhuys2023}. 
In the latter case the collective anisotropic vortex/antivortex states have been predicted \cite{Fistul2020} and studied in detail \cite{Neyenhuys2023}. In particular, it has been shown in Ref. \cite{Neyenhuys2023} that the physical origin of these collective states is the presence of a huge amount of topological constraints caused by the flux quantization in closed superconducting loops. This induces a highly anisotropic interaction between Josephson junctions of different unit cells. At the same time, in Ref. \cite{Fistul2019} it has been shown that in quasi-1D \textit{f}-JJAs, e.g., saw-tooth and diamond chains of Josephson junctions, the interaction between Josephson junctions of different cells is absent resulting in random configurations of vortices/antivortices in the classical frustrated regime.

In the absence of interaction, the quantum dynamics of quasi-1D vertex-sharing \textit{f}-JJAs is reduced to a set of \textit{independent} flux qubits, and there are no collective quantum phases. Therefore, natural questions arise in this field: which interaction allows to observe various collective quantum phases in quasi-1D \textit{f}-JJAs and how it can be realized?

In this paper, we answer these questions demonstrating that a straightforward embedding of a frustrated saw-tooth chain of Josephson junctions in a dissipationless transmission line allows to establish a strong, intrinsically quantum interaction between Josephson junctions of different cells. Moreover, depending on the physical parameters this interaction is either long- or short-ranged. We show how various collective quantum phases and quantum phase transitions between them can be observed in quasi-1D $f$-JJAs. 

The paper is organized as follows: In Section~\ref{sec2:level1} we introduce the electrodynamic model of frustrated saw-tooth quasi-1D arrays of small (\textit{quantum}) Josephson junctions incorporated in a dissipationless transmission line and define the most important physical parameters and dynamic variables of the system. We write down the potential and kinetic energies, the Lagrangian and the  Hamiltonian for two Josephson junction arrays: a frustrated single triangular cell of Josephson junctions (Sec. \ref{sec2:level2}) and a long frustrated saw-tooth chain of Josephson junctions (Sec. \ref{sec2:level3}). In the latter case we obtain a long-range charge interaction between Josephson junctions of well separated cells. 
In Section \ref{sec3:level1}, we address the frustrated regime in which the potential energy of a single cell has two equivalent minima. In Section \ref{sec3:level2} the variational approach is used to reduce the circuit Hamiltonian to an effective spin Hamiltonian. In the Section \ref{sec3:level3}, we illustrate this method for a frustrated single triangular cell of Josephson junctions. In Section \ref{sec4:level1}, the quantum dynamics of low-lying eigenstates will be mapped to an effective $XX$ quantum spin chain with a long/short-range exchange interaction and a local magnetic field applied in the $x$-direction. The collective quantum phases and corresponding quantum phase transitions will be identified in Section \ref{sec5:level1}. Section~\ref{sec6:level1} provides conclusions. 

\section{Electrodynamic Model, Dynamic Variables, Lagrangian and Hamiltonian} 
\label{sec2:level1}
Let us consider an exemplary frustrated vertex-sharing saw-tooth array of Josephson junctions. Such an array contains $N$ periodically arranged basic cells, i.e., triangular superconducting loops interrupted by two $0$- and one $\pi$-Josephson junctions. The $0$- and $\pi$-Josephson junctions connecting adjacent superconducting islands, i.e., nodes of the array, are characterized by the Josephson energies $E_J = \hbar I_c/(2e)$ and $\alpha E_J$, and the charging energies $E_C= e^2/(2C)$ and $E_C/|\alpha|$, respectively. Here, $I_c$ and $C$ are the critical current and the capacitance of the $0$-Josephson junction, respectively. The parameter $\alpha$ varying between $-1$ and $1$ determines the \textit{frustration} of a system and is related to the commonly used frustration parameter $f$ as $\alpha = (1-2f)$, where $0<f<1$. 
Utilizing the $\pi$-Josephson junction with the parameter $\alpha<\alpha_c=-0.5$ ($f>f_{c}=3/4$), one can realize the frustrated regime in which each basic cell of the array is in one of two equivalent classical ground states. In such $f$-JJAs, the interaction between Josephson junctions of different cells is absent, and the classical ground state of the entire array is $2^N$ degenerate.

A natural way to provide a strong intrinsically quantum interaction in vertex-sharing quasi-1D JJAs is to embed the $\pi$-Josephson junctions as distributed inductances in a dissipationless transmission line. The transmission line is also characterized by the capacitance to the ground, $C_0$. Although in most of the experiments so far, the ratio of $C_0/C$ was rather small \cite{Haviland2000, masluk2012microwave}, the opposite limit $C_0 \geq C$ was also obtained in specially prepared Josephson metamaterials, e.g., SQUID transmission lines \cite{Ranadive2022, planat2019fabrication}. The schematic of a frustrated saw-tooth chain of Josephson junctions incorporated in the transmission line is presented in Fig. \ref{fig:one}.

The classical electrodynamics of a system is completely determined by the \emph{two} time-dependent phases of the superconducting order parameter per unit cell, $\chi_n= \{\chi_{0,n}; \chi_{+,n} \}$. By making use of the Kirchhoff's laws and choosing a spanning tree (graph theory) as introduced in Refs. \cite{Devoret2017, Rasmussen2021}, we obtain the Lagrangian $L=K-U$, where the potential $U$ and the kinetic $K$ energies are written as follows 
\begin{align}
    \label{eq:PotentialEnergy}
    U(\{ \chi_n\}) &=  E_J\sum_{n=1}^{N}[2+\alpha - \cos( \chi_{+,n} - \chi_{0,n})  \notag \\
   - & \cos(\chi_{0,n+1} - \chi_{+,n}) - \alpha \cos( \chi_{0,n} - \chi_{0,n+1}) ]
\end{align}
and 
\begin{align}
    \label{eq:KineticEnergy}
    K(\{\dot \chi_n\}) &= \frac{\hbar^2}{16 E_C}\sum_{n=1}^{N} [( \dot \chi_{+,n} - \dot \chi_{0,n})^2 + (\dot \chi_{0,n+1} - \dot \chi_{+,n})^2 \notag \\
    & +|\alpha|(\dot \chi_{0,n} - \dot \chi_{0,n+1})^2 ] +\frac{\hbar^2}{16E_{C_0}}\sum_{n=1}^{N+1} (\dot \chi_{0,n})^2,
\end{align}
where the charging energy to the ground $E_{C_0}=(C/C_0)E_C$.

\subsection{\label{sec2:level2}Effective Hamiltonian of a single building block}

\begin{figure}[b]
    \centering
    \includegraphics[width=0.35\textwidth]{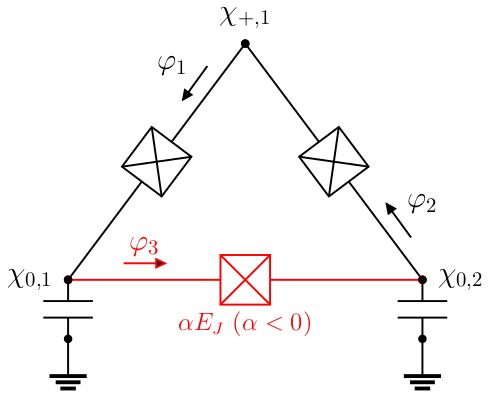}       
\caption{\label{fig:2} (color online) The schematic of a single building block of frustrated saw-tooth chains of Josephson junctions. The phases of the order parameter of superconducting islands, $\chi_0$ ($\chi_+$), and corresponding Josephson phases $\varphi$ are shown.} 
\end{figure}

A frustrated saw-tooth chain of Josephson junctions consists of periodically arranged building blocks, i.e.,  triangular superconducting loops interrupted by three (two -$0$ and one -$\pi$) Josephson junctions. The schematic of a single building block is presented in Fig. \ref{fig:2}. 
To elaborate the classical and quantum dynamics of a single building block we introduce the Josephson phases as $\varphi_1=\chi_{+,1}-\chi_{0,1}$, $\varphi_2=\chi_{0,2}-\chi_{+,1}$ and $\varphi_3=\chi_{0,1}-\chi_{0,2}$. For superconducting loops of a small area the Josephson phases satisfy the flux quantization condition, $\varphi_1+\varphi_2+\varphi_3=0$. 
As we are interested in the low-energy dynamics of the system, we make a few further simplifications. First, it is not necessary to take into account the $2\pi$-periodicity of the potential energy landscape, because the energy barriers $\simeq E_J$ are assumed to be large compared to both $k_B T$, where $T$ is the temperature, and $\hbar \Omega$, where $\Omega$ is a typical frequency of small oscillations. 

\begin{figure}
    \centering
    \includegraphics[width=.4\textwidth]{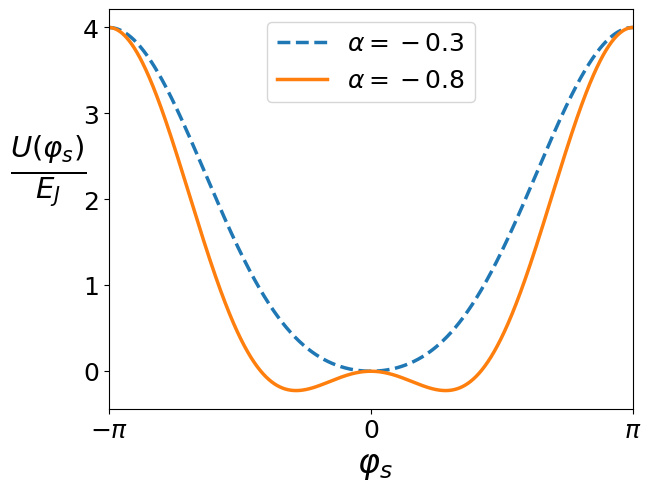}
    \caption{(color online) Calculated effective potential energy of a single building block. In the non-frustrated regime there is a single minimum at $\varphi_s = 0$ (dashed line), and in the frustrated regime there are two shallow minima $\varphi_s = \pm u_0 $ (solid line).  \label{fig:double_well}}
\end{figure}

Taking into account the flux quantization condition, one finds that the potential energy is determined by two Josephson phases $\varphi_{1,2}$. It is convenient to symmetrize these variables by introducing the symmetric and antisymmetric combinations $\varphi_{s,a} = (\varphi_1 \pm \varphi_2)/2$. In the frustrated regime $-1<\alpha <-0.5$, the potential energy has two equivalent shallow minima at $\varphi_s = \pm u_0$ with $u_0 = \arccos[1/(2|\alpha|)]$ and $\varphi_a = 0$ (see Fig. \ref{fig:double_well}). These minima are separated by the potential energy barrier $E_J(\alpha)=-E_J[2(1+\alpha)+1/(2\alpha)]$, which becomes zero at the critical value of $\alpha=\alpha_c=-0.5$. Since the frequency of small oscillations in the $\varphi_a$ direction is much larger than those in the $\varphi_{s}$ direction, they cannot be excited at low temperatures. Therefore, the potential energy of a single building block of frustrated saw-tooth chains of Josephson junctions is effectively determined by a single dynamic variable, $\varphi_s$. 

In the classical frustrated regime, the two minima of the potential energy correspond to non-zero persistent currents flowing in two opposite (clockwise or anticlockwise) directions. Note that the frustrated regime, as defined here, is equivalent to the so-called \textit{flux qubit} biased to the symmetry point \cite{Orlando1999}, i.e., an externally applied magnetic flux $\Phi=(1/2) \Phi_0$, where $\Phi_0$ is the flux quantum. 
The kinetic energy (\ref{eq:KineticEnergy}) of a single building block depends on $\dot \varphi_s$ and $\dot \chi_{0,1}$. The latter determines the dynamics of the whole system and does not affect the collective states of the system. Thus, we further disregard this dynamics and set $\dot \chi_{0,1}$ to zero. 

With these simplifications, we can now express the circuit Lagrangian for a single building block of a system as (from here, we will omit the index $s$ in $\varphi_s$),
\begin{eqnarray}
    \mathcal{L}_{sb} && =  \frac{\hbar^2}{8 E_C}\left(\gamma + 1 + 2|\alpha| \right) \dot{\varphi}^2 \nonumber \\ 
    && - E_J \left [2+\alpha -2 \cos(\varphi) - \alpha \cos(2\varphi) \right],
\end{eqnarray}
where the parameter $\gamma = C_0/C$ is introduced. The characteristic frequency $\Omega$ of small oscillations of the Josephson phase around the minima of the potential energy is 
\begin{eqnarray}
\Omega=[2/(\hbar)] \sqrt{E_CE_J(|4\alpha+1/|\alpha||)/(\gamma+1 +2|\alpha|)}. 
\end{eqnarray}
We also define the effective mass: 
\begin{eqnarray}
m_{eff}=(\hbar^2/4E_C)(\gamma+1 +2|\alpha|).
\end{eqnarray}

Introducing the generalized momentum $p=\partial \mathcal{L}_{sb}/\partial \dot \varphi$ and representing it in the operator form $\hat p=-i\hbar \partial /\partial \varphi$ we arrive at the circuit Hamiltonian
\begin{eqnarray}
    \hat H_{sb}&& = \frac{2 E_C}{\hbar^2} \frac{1}{\gamma + 1 + 2|\alpha|} \hat p^2 \nonumber \\ 
    && + E_J \left [2+\alpha -2 \cos(\varphi) - \alpha \cos(2\varphi) \right].
    \label{eq:H_flux}
\end{eqnarray}

\subsection{\label{sec2:level3} Effective Hamiltonian of frustrated saw-tooth chains of Josephson junctions}

Using the assumptions discussed in Sec. \ref{sec2:level2} and introducing the Josephson phases as $\vec \varphi =\{\varphi_n\}$, where $\varphi_n=\chi_{+,n}-\chi_{0,n}$, and $n$ is the cell number, the Lagrangian of a saw-tooth chain of Josephson junctions embedded in the transmission line (see Fig. \ref{fig:one}) is written as follows
\begin{eqnarray}
    \mathcal{L} && =  \frac{\hbar^2}{8 E_C}   \left[  (1 + 2|\alpha|) \dot{\vec{\varphi}}^T\dot{\vec{\varphi}} + \dot{\vec{\varphi}}^T \mathbf{C}_{\gamma} \dot{\vec{\varphi}} \right] \nonumber \\ 
    && - E_J \sum_i \left( 2 + \alpha-2 \cos(\varphi_i) - \alpha \cos(2\varphi_i) \right), 
\end{eqnarray}
where the non-diagonal capacitance matrix $\mathbf{C}_{\gamma}$ determines the \textit{interaction} between the Josephson junctions of different cells and is expressed explicitly as
\begin{eqnarray}
    \mathbf{C}_{\gamma} = 2 \gamma \begin{bmatrix}
        N & N-1 & N-2 & \ldots & 1 \\ 
        N-1 & N-1 & N-2 & \ldots & 1 \\ 
        N-2 & N-2 & N-2 & \ldots & 1 \\ 
        \vdots & \vdots & \vdots & \ddots & 1 \\ 
        1 & 1 & 1 & 1 & 1 
    \end{bmatrix}.
\end{eqnarray}
Similar to Sec. \ref{sec2:level2}, the generalized momenta are defined as $\vec p=\partial \mathcal{L}/\partial \dot{\vec \varphi}$. Applying the Fourier transform $\vec \varphi \rightarrow \vec \varphi_k$ and utilizing the special tri-diagonal form of the inverse capacitance matrix $\mathbf{C}_{\gamma}^{-1}$, we derive the circuit Hamiltonian in the $k$-space as $\hat{H}_k = \frac{2E_C}{\hbar^2} \sum_k \Gamma(k) \hat p_k \hat p_{-k} + \sum_k U(\phi_k)$, where the non-local kinetic energy factor $\Gamma (k)$ is
\begin{eqnarray}
    \Gamma(k) =  \frac{\sin^2(k/2)}{\gamma/2 + (1 + 2|\alpha|) \sin^2(k/2)}.
    \label{eq:Gammak}
\end{eqnarray}
Performing the inverse Fourier transform, the effective circuit Hamiltonian is then written as
\begin{eqnarray}
    \hat{H} && =  \frac{2 E_C}{\hbar^2}  \left[ \Gamma_d \sum_i  \hat p_i^2 - \Gamma_{o} \sum_{i\neq j} e^{- |i-j|/\ell_0} \hat p_i \hat p_j \right] \nonumber \\ 
    && + E_J \sum_i \left [ 2 + \alpha -2 \cos(\varphi_i) - \alpha \cos(2\varphi_i) \right] 
    \label{eq:Hsawt-3}
\end{eqnarray}
where the parameters $\Gamma_d = \frac{1}{2|\alpha| +1} - \frac{\gamma}{(2|\alpha| +1) \sqrt{\gamma (4|\alpha| + \gamma + 2)} } $ and $\Gamma_o = \frac{\gamma}{(2|\alpha| +1) \sqrt{\gamma (4|\alpha| + \gamma + 2)} }$, and the interaction length,  $\ell_0 = \left| \log( \frac{2|\alpha| + 1 + \gamma - \sqrt{\gamma (4|\alpha| + \gamma + 2) }}{2|\alpha| + 1 }) \right|^{-1}$. 

\begin{figure}
    \centering
    \includegraphics[width=.4\textwidth]{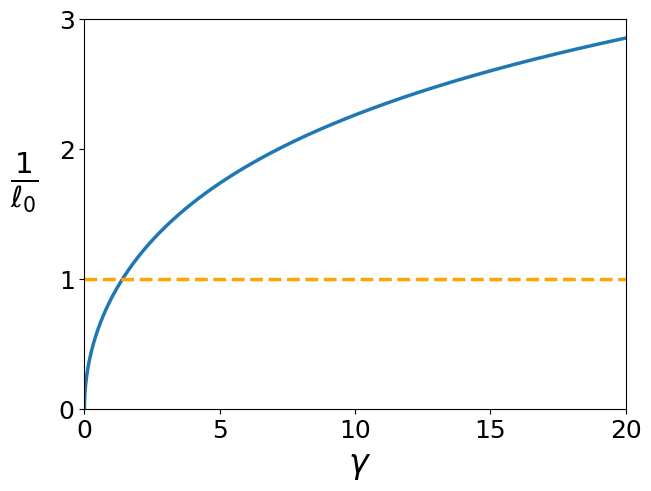}
    \caption{(color online) Calculated dependence of the inverse interaction length $1/\ell_0$ on the ratio of the ground $C_0$ and the Josephson junction $C$ capacitances, $\gamma=C_0/C$. We choose $\alpha=-0.8$. 
    }
    \label{fig:int_range}
\end{figure}

The term proportional to $\Gamma_o$ determines the interaction between the cells. Therefore, if $\gamma=0$ the quantum dynamics of a saw-tooth chain of Josephson junctions is reduced to the dynamics of independent cells. As $\gamma = C_0/C \ll 1$, the interaction strength is small but extends over a long range, $\ell_0 \simeq 1/\sqrt{\gamma} $.
In the limit of $\gamma \gg 1$, only the nearest neighbor interactions remain. The typical dependence of the inverse interaction length $1/\ell_0$ on the parameter $\gamma$ is presented in Fig. \ref{fig:int_range}.
To conclude this section, we notice that the interaction length $\ell_0$ has a physical meaning of the charge screening length \cite{Haviland2000}. 

\section{\label{sec3:level1}Effective quantum spin model}

In the frustration regime, $\alpha<\alpha_{c}=-0.5$ ($f>f_{c}=3/4$), the potential energy of a system exhibits $2^N$ equivalent minima separated by small potential barriers (see the solid line in Fig. \ref{fig:double_well} for a single building block). At low temperatures, as $k_B T \ll \hbar \Omega$, where $\Omega$ is the characteristic frequency around each minimum,  the coherent quantum regime is established, and in this regime the quantum dynamics of a system consists of small oscillations around the positions of the minima and quantum tunneling between them. The latter corresponds to the coherent quantum beats between persistent currents of opposite directions. Equivalently, it can be understood as the coherent quantum beats between vortex/antivortex penetrating each cell. The amplitude of such macroscopic quantum tunneling is exponentially small if the parameter $\alpha$ is not too close to the critical value $\alpha_{c}$.  
To analyze the coherent quantum frustrated regime, we use the variational approach allowing to reduce the circuit Hamiltonian (\ref{eq:Hsawt-3}) to \textit{an effective spin Hamiltonian}. In this section, we apply this method to a single building block of a frustrated saw-tooth Josephson junction array (see Fig. \ref{fig:2}).

\subsection{\label{sec3:level2}Variational approach}
Here, we present an outline of the procedure. We choose the wave function of a whole system as the product of the localized wave functions taking a Gaussian form around each minimum
\begin{equation}
    \Psi(\vec{\varphi}, \vec{\sigma}^{(n)}_{z}) = \frac{1}{R} \exp \left [- \left ( \vec{\varphi} - u_0 \vec{\sigma}^{(n)}_{z}\right) \mathbf{A} \left( \vec{\varphi} - u_0 \vec{\sigma}^{(n)}_{z} \right) \right ],
    \label{eq:trial_wf}
\end{equation}
where $1/R$ is the normalization factor. The presence of two minima in each cell is denoted by the classical spin component, $\sigma_z=\pm 1$. Thus, the wave function $\Psi_n$ is determined by the vector of Josephson phases, $\vec{\varphi}=\{\varphi_i \}$, and the $n$-th configuration of the spin vector, $\vec{\sigma}^{(n)}_{z}=\{\sigma_{z,i}\}$, where $i$ is the cell number. 
Using these functions as a basis, the matrix elements of an effective spin Hamiltonian are calculated as 
\begin{equation} \label{SG-matrixelements}
    \mathbf{H}_{nl} = \int d\vec{\varphi} \Psi(\vec{\varphi}, \vec{\sigma}^{(n)}_{z}) \hat H  \Psi(\vec{\varphi}, \vec{\sigma}^{(l)}_{z}).
\end{equation}
For Josephson arrays with $N$ cells, we have $2^N$ basis functions, and $\mathbf{H}$ is a $2^N\times 2^N$ matrix. 

The $N \times N$ matrix $\mathbf{A}$ is obtained by minimizing the ground state expectation value, $E_0 = \mathbf{H}_{nn} = \int d\vec{\varphi} \Psi(\vec{\varphi}, \vec{\sigma}^{(n)}_{z}) \hat {H} \Psi(\vec{\varphi}, \vec{\sigma}^{(n)}_{z}) $. Other matrix elements of
the effective spin Hamiltonian $\mathbf{\hat H}$ contain exponentially small factors $G_{n,l}$ obtained as 
\begin{equation}
   \mathbf{H}_{nl} \propto  G_{n,l} = \exp \left [ - \frac{1}{2} u_0^2 \left( \vec{\sigma}^{(n)}_{z} -  \vec{\sigma}^{(l)}_{z} \right) \mathbf{A}  \left( \vec{\sigma}^{(n)}_{z} -  \vec{\sigma}^{(l)}_{z} \right)  \right ]. 
    \label{eq:Gnl}
\end{equation}
Next, we define $\vec{Z}_{n,l} = \left( \vec{\sigma}^{(n)}_{z} -  \vec{\sigma}^{(l)}_{z} \right)$. This vector is zero if no quantum tunneling events are present upon going from the configuration $(n)$ to $(l)$ or vice versa. A single tunneling event in the $j$th-cell leads to a non-zero element $Z_j = \pm 2$, and the term $\propto \hat \sigma_{x,j}$ in the effective spin Hamiltonian $\mathbf{\hat H}$. In this framework, a single tunneling event is equivalent to a \textit{spin-flip} of the $j$th spin. Multiple simultaneous tunneling events, i.e., spin-flips occurring in different cells, are interpreted as an interaction between different spins.

\subsection{\label{sec3:level3} Effective spin Hamiltonian: a single building block}
For the single building block of a frustrated Josephson junction array, an application of the procedure elaborated above results in the effective spin Hamiltonian:
\begin{eqnarray} \label{SH-sb}
  \mathbf{\hat H}_{sb}=\Delta_{sb} \hat \sigma_x.
\end{eqnarray}
The quantum tunneling amplitude, i.e., a single spin-flip, is obtained as follows: the matrix $\mathbf{A}$ is reduced to a single parameter $A$, and utilizing harmonic approximation for the potential $U(\varphi)$ we get the analytical result, $A=m_{eff}\Omega/(2\hbar)$, or explicitly $A=\sqrt{(E_J/16 E_C)(|4\alpha+1/|\alpha||)(\gamma+1 +2|\alpha|)}$. Calculating the Gaussian integrals in Eq. (\ref{SG-matrixelements}) we obtain the non-diagonal matrix element of Eq. (\ref{SH-sb}) containing a small exponential factor as
\begin{eqnarray} \label{MElement-SB}
\Delta_{sb} \propto G_1=\exp(- 2 u_0^2 A ).
\end{eqnarray}
The pre-exponential factor in Eq. (\ref{MElement-SB}) was calculated numerically, and was obtained to be of the order $\hbar \Omega$.

Additionally, one can find the quantum tunneling amplitude by making use of the quasiclassical ($WKB$) approximation \cite{Landau-Lifshitz}.
In a complete analogy with Refs. \cite{Orlando1999,Landau-Lifshitz} the tiny energy gap $2\Delta_{sb}$ between the two lowest eigenvalues is written as: 
\begin{eqnarray}\label{WKB}
\Delta^{WKB}_{sb} = \frac{\hbar \Omega}{2\pi} \exp \left(-\frac{2\sqrt{2m_{eff}E_J}}{\hbar}\left( \sqrt{2|\alpha| - \frac{1}{2|\alpha|}} \right. \right. \nonumber \\
\left. \left. - \frac{1}{\sqrt{2|\alpha|}} \arccos( \frac{1}{2|\alpha|}) \right)\right). ~~~~~~~~~~~~~~~~~~~~~ 
\end{eqnarray}
In order to compare the results for $\Delta_{sb}$ obtained by two methods, one can study small exponential factors in the limit $\alpha \rightarrow \alpha_{c}$. In this limit, the variational approach results in $\ln G_1=- \sqrt{\frac{E_J}{E_C}(\gamma + 2)} \left(2|\alpha|-1 \right)^{3/2}$, while using the quasiclassical approximation one obtains $\ln\left(\frac{2\pi\Delta^{WKB}_{sb}}{\hbar \Omega}\right) = - \frac{4}{3} \sqrt{\frac{E_J}{E_C}(\gamma+2)} \left( 2|\alpha|-1 \right)^{3/2}$. Therefore, these two methods demonstrate the same universal dependence of a small exponential factor in Eqs. (\ref{MElement-SB}) and (\ref{WKB}) on the parameter $\alpha$ as $\alpha \simeq \alpha_{c}$, with only different numerical factors. 

Therefore, one can expect that the variational approach can be effectively used to describe low-energy quantum dynamics of long frustrated Josephson junction arrays.  

\section{\label{sec4:level1}Effective $XX$ spin Hamiltonian with a long-range interaction}

Here, we apply the variational approach, elaborated in the Section \ref{sec3:level1}, to derive an effective spin Hamiltonian for a long frustrated saw-tooth chain of Josephson junctions. Using the trial wave function, Eq. (\ref{eq:trial_wf}), and calculating the expectation value of the ground state energy, $E_0 (\mathbf{A})= \mathbf{H}_{nn} =\int d\vec{\varphi} \Psi(\vec{\varphi}, \vec{\sigma}^{(n)}_{z}) \hat {H} \Psi(\vec{\varphi}, \vec{\sigma}^{(n)}_{z}) $, we obtain the matrix $\mathbf{A}$ by minimizing $E_0$. Due to the specific structure of the interaction in the Hamiltonian (Eq. (\ref{eq:Hsawt-3})), it is convenient to rewrite the Hamiltonian in the $k$-space and use the harmonic approximation:
\begin{eqnarray}
    \hat{H}_{\textrm{harm}} = \frac{1}{2 m_k} \hat p_k \hat p_{-k} + \frac{m_k\Omega_k^2}{2} \varphi_k \varphi_{-k}, 
\end{eqnarray}
where we identify $m_k = \frac{\hbar^2}{4E_C} \frac{1}{\Gamma(k)}$ and $\Omega_k = \sqrt{\frac{E_J}{m_k}(4|\alpha| - 1/|\alpha|) }$. In the $k$-space, the trial wave function has a simple form: $\Psi_k = 1/R_k \exp(-\zeta_k \phi_k^2)$. It allows us to obtain the parameter $\zeta_k$ as 
\begin{eqnarray}
    \zeta_k = \frac{m_k \omega_k}{2\hbar} = \frac{D}{4}  \frac{\sqrt{\gamma/2 + (2|\alpha|+ 1) \sin^2(k/2)}}{|\sin(k/2)|},
\end{eqnarray}
where $D = \sqrt{ \frac{E_J}{E_C}\left( 4|\alpha| - \frac{1}{|\alpha|}\right) }$. Making the inverse Fourier transformation to the real space, we obtain the dependence of the matrix elements $A_{nm}$ on the distance between $n$-th and $m$-th cells:
\begin{eqnarray}
    A(|n-m|) = 2 \sum_k \zeta_k \cos[k(n-m)].
\end{eqnarray}
In this analysis we imply open boundary conditions. 

The analysis of the non-diagonal matrix elements of the effective spin Hamiltonian $\mathbf{\hat H}$ (Eq. (\ref{eq:Gnl})) allows us to conclude that multiple simultaneous spin flips involving $n$ cells with $n>2$ are exponentially small compared to the spin flips with $n=1$ and $n=2$ (see the Appendix for details). Taking this property into account, we write the effective spin Hamiltonian in the following form:
\begin{equation} \label{eq:EffSpinHam-Long}
\mathbf{\hat H}=\sum_{n=1}^N \Delta \hat \sigma_{x,n}+\frac{1}{2}\sum_{n \neq m} J(|n-m|)\left [\hat \sigma_{x,n}\hat \sigma_{x,m}+\hat \sigma_{y,n}\hat \sigma_{y,m}\right].
\end{equation}
Here, the first and second terms in the r.h.s. of Eq. (\ref{eq:EffSpinHam-Long}) describe the quantum tunneling between a vortex and an antivortex in a single cell, and induced by the coupling to a transmission line the long/short-range exchange spins interaction between different cells, accordingly. The amplitude of quantum tunneling, $\Delta$, and the exchange interaction strength, $J(|n-m|)$, are determined by the matrix elements $A_{nm}$ as
\begin{eqnarray} \label{Singleflip-longJJA}
\Delta =\Delta_0 \exp[- 2 u_0^2 A(0)]
\end{eqnarray}
and 
\begin{equation} \label{Twospinsflip-longJJA}
J(|n-m|) = J_0\exp  \{- 4 u_0^2 [A(0)-A(|n-m|)] \},
\end{equation}
where both pre-exponential factors, $\Delta_0$ and $J_0$, are of the order of $\hbar \Omega$.

\subsection{\label{sec4:level2}  The parameters $\Delta$ and $J(|n-m|)$: general properties}

The parameters $\Delta$ and $J(|n-m|)$ of the effective spin Hamiltonian $\mathbf{\hat H}$ are determined by \textit{two} matrix elements $A(0)$ and 
$A(|n-m|)$ having the following properties: in the absence of the coupling to the transmission line, i.e., as  $C_0 = 0$, the matrix $A(|n-m|)$ has a diagonal form, $A(|n-m|)=(m_{eff}\Omega/\hbar)\delta_{nm}$. Therefore, the exchange interaction term in the Hamiltonian (\ref{eq:EffSpinHam-Long}) is absent. In this case the value of $\Delta$ coincides with $\Delta_{sb}$ (Eqs. (\ref{MElement-SB}) and (\ref{WKB}) for $C_0=0$).

In a general case of $C_0 \neq 0$, the matrix elements $A(0)$ and $A(|n-m|)$ strongly depend on the ratio of the total length of the array $N$ to the interaction length $\ell_0$. For short arrays, as $N< \ell_0$, we obtain that the exchange interaction term in the Hamiltonian (\ref{eq:EffSpinHam-Long}) is small and weakly decays with the distance $|n-m|$. The parameter $\Delta \simeq \Delta_{sb}$. 

\begin{figure}
    \centering
    \includegraphics[width=.49\textwidth]{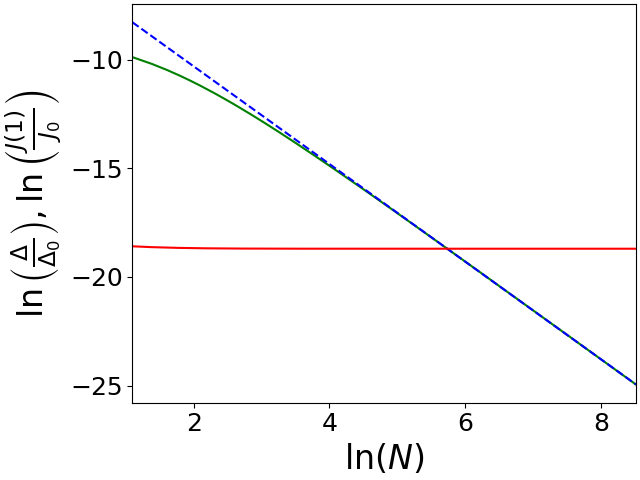}
    \caption{(color online) Calculated dependencies of the amplitude of quantum tunneling, $\Delta$ (green solid line), and the exchange interaction strength, $J(1)$ (red solid line), on the total length $N$. The blue dashed line indicates the logarithmic behavior of $\ln(\Delta/\Delta_0)$ as $N>\ell_0$. The chosen parameters are $\alpha = -0.8$, $\gamma = 1.0$, and $E_J/E_C = 80$.
    }
    \label{fig:log_div_Delta}
\end{figure}

However, the most interesting case is realized for long arrays as $N \gg \ell_0$. Here, the parameter $A(0)$ increases logarithmically with $N$, i.e., $A(0) = (D/2)\sqrt{\gamma/2} \ln (N/\ell_0)$, and therefore, the amplitude of the quantum tunneling between vortex and antivortex in a single cell is strongly suppressed. The typical dependence of $\Delta(N)$ is shown in Fig. \ref{fig:log_div_Delta} (green solid line). On the other hand, the parameter $[A(0)-A(1)]$, determining the strength of the exchange interaction $J(1)$, weakly depends on $N$ (see, red line in Fig. \ref{fig:log_div_Delta}). This means that varying the length of the array $N$ or the interaction length $\ell_0(\gamma)$ enables one to switch between strong and weak exchange interactions. 

\begin{figure}
    \label{fig:b_ga_and_d_b_approx}
    \subfloat[]{
    \includegraphics[width=0.95\columnwidth]{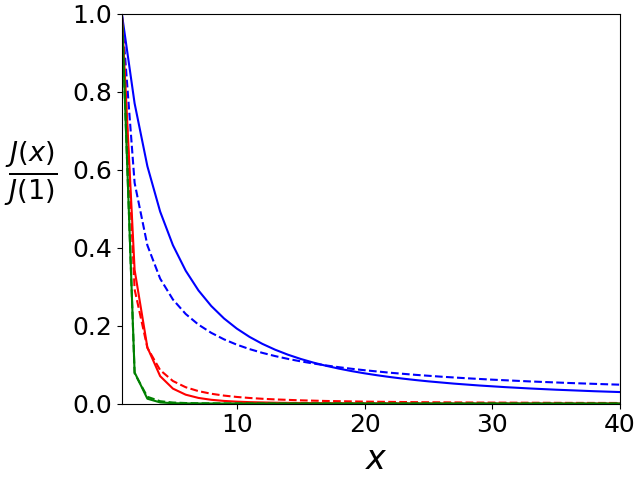}
    \label{fig:log_div_J_ij}}
    \\
    \vspace{-0.8\baselineskip}
    \subfloat[]{
    \includegraphics[width=0.95\columnwidth]{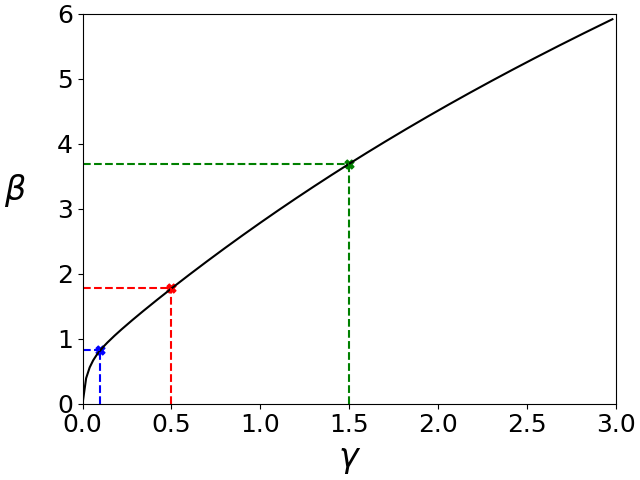}
    \label{fig:beta_gamma}}
    \caption{(color online) (a): Calculated typical dependencies of $J(|x|)/J(1)$ (solid lines) for different values of $\gamma$: $\gamma=1.5$ (green line), $\gamma=0.5$ (red line) and $\gamma=0.1$ (blue line). The approximations with the power-law dependence are shown by colored dashed lines. Other parameters were chosen as $\alpha=-0.8$, $N=1000$, $E_J/E_C=80$. (b): The calculated dependence of the power exponent $\beta$ on the parameter $\gamma$. The colored dashed lines correspond to the $J(|x|)$ curves presented in Fig. \ref{fig:log_div_J_ij}. Other parameters were chosen as $\alpha=-0.8$, $N=1000$, $E_J/E_C=80$.  }
\end{figure}

At the distance $|n-m|> \ell_0$, the parameter $A(|n-m|)$ increases with $|n-m|$ as $A(|n-m|) = (D/2) \sqrt{\gamma/2} \times \ln (N/|n-m|)$. Therefore, the exchange interaction term can be accurately approximated by a power-law $J(x) \approx J(1) (|x|)^{-\beta}$ with
$\beta = (Du_0^2)\sqrt{\gamma/2}$. 
The typical dependencies of $J(|x|)$, together with the power-law approximations (shown by dashed lines) for different values of the parameter $\gamma$, are presented in Fig. \ref{fig:log_div_J_ij}. The obtained dependence of the power exponent $\beta$ on the parameter $\gamma$ is presented in Fig. \ref{fig:beta_gamma}.

Thus, varying the parameters $E_J$, $E_c$, $\alpha$ and $\gamma$, one can realize different regimes of the effective spin Hamiltonian (\ref{eq:EffSpinHam-Long}): long- or short-range exchange spin interactions, as well as strong/weak local quantum tunneling amplitude.        

\section{\label{sec5:level1} Collective phases of the effective spin Hamiltonian}
Here, we present a numerical analysis of the effective $XX$ spin Hamiltonian in the situation  when the exchange spin-spin interaction decays with distance according to a power law. We rewrite Eq. (\ref{eq:EffSpinHam-Long}) in the dimensionless form:
\begin{equation} \label{eq:EffSpinHam-Long2}
\frac{\mathbf{\hat H}}{J(1)}=\sum_{n=1}^N \tilde{\Delta}\hat \sigma_{x,n}+\frac{1}{2}\sum_{n \neq m} \frac{1}{|n-m|^\beta}\left [\hat \sigma_{x,n}\hat \sigma_{x,m}+\hat \sigma_{y,n}\hat \sigma_{y,m}\right],
\end{equation}
where we introduce the dimensionless parameter $\tilde \Delta=\Delta/J(1)$.

This Hamiltonian is not integrable (except the case of $\Delta=0$) \cite{Ovchinnikov2002,Ovchinnikov_2_2002}, and therefore, our further analysis is based on the direct numerical diagonalization of Eq. (\ref{eq:EffSpinHam-Long2}) for spin chains of a moderate size, up to $N=16$. Exploiting open boundary conditions and fixing the parameters $\tilde \Delta $, $\beta$, we obtain the eigenvalues $E_\alpha$ and eigenvectors $\ket{\psi_\alpha}$, where $\alpha$ is the label of the different energy levels. Instead of performing the complete numerical diagonalization \cite{Sandvik2010}, we use Arnoldi method \cite{Arnoldi1951} to find the eigenvalues and eigenvectors of the ground and first excited state $E_0$, $E_1$, $\ket{\psi_0}$ and $\ket{\psi_1}$. Next, we vary the parameters $|\tilde{\Delta}|<15$ and $0 <\beta<5$, to explore the different collective phases. 

\begin{figure}
    \subfloat[]{
    \includegraphics[width=0.95\columnwidth]{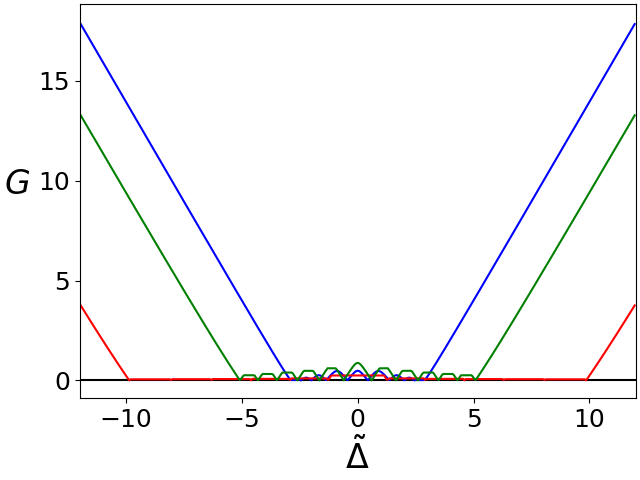}
    \label{fig:ED_gap_cut}
    }
    \\
    \vspace{-0.8\baselineskip}
    \subfloat[]{
    \includegraphics[width=0.95\columnwidth]{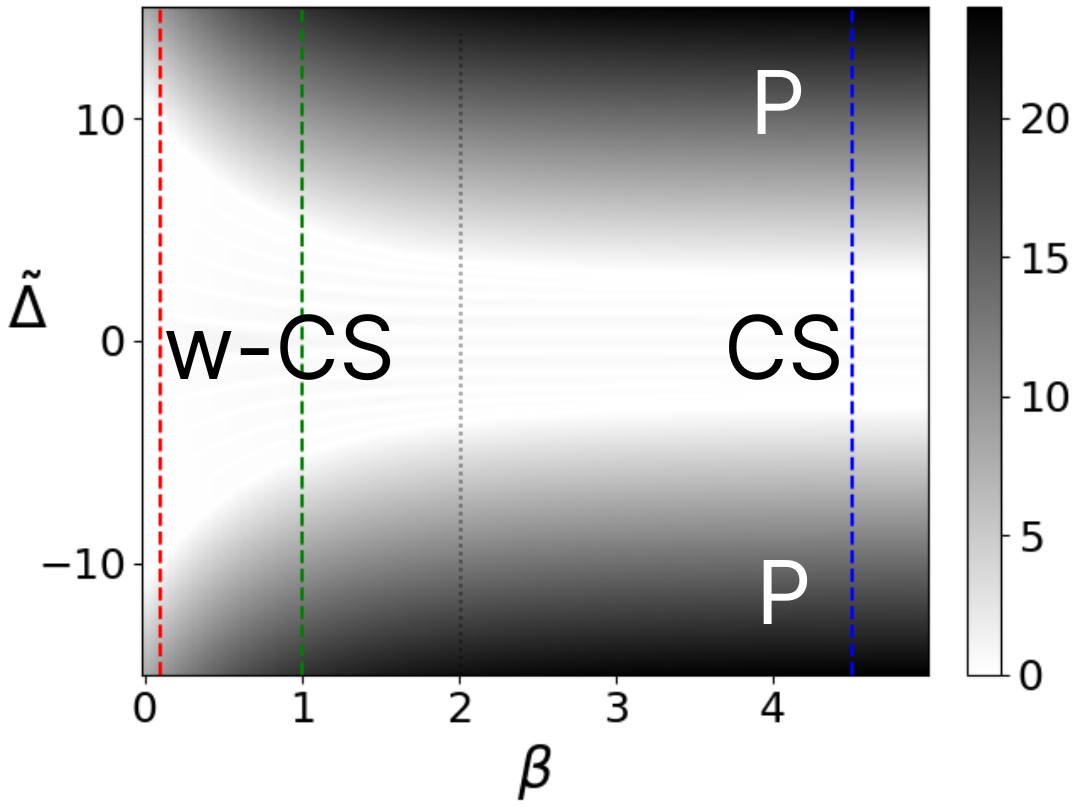}
    \label{fig:ED_gap_2d}
    }
    \caption{ (color online)(a) Calculated dependence of the minimal energy gap, $G=(E_1-E_0)/J(1)$, on the ratio of the local tunneling amplitude to the maximum exchange interaction strength, $\tilde \Delta=\Delta/J(1)$ for various values of $\beta$: $\beta = 4.5$ (blue line), $\beta = 1.0$ (green line), $\beta = 0.1$ (red line); 
    (b) The color plot showing the dependence of the minimal gap $G$ on $\tilde \Delta$ and $\beta$. The color dashed lines corresponds to the values of $\beta$ presented in Fig. \ref{fig:ED_gap_cut}. 
    The other physical parameters were chosen as  $N=12, \alpha = -0.8, E_J/E_C = 80$. }
\end{figure}

We characterize the collective quantum phases by the dimensionless minimum energy gap, i.e., $G=(E_1-E_0)/J(1)$. In Fig. \ref{fig:ED_gap_cut}, the dependencies of the minimum energy gap $G$ on the relative amplitude of the $\sigma_x$ term, $\tilde \Delta$, for different values of $\beta$ are presented. One can see that for the large values of $|\tilde \Delta|$, the interaction term becomes relatively small, and extensive local spin-flips lead to the \textit{paramagnetic} ($P$) state. 
The $P$-state is characterized by a finite value of the minimum energy gap $G \propto |\tilde \Delta|$, a low entanglement, and the quantum-mechanical average of the total $x$-component of the magnetization, $M_x=1/N <\sum_i \sigma_{x,i}>$ is one. 

As the parameter $|\tilde \Delta|$ decreases below the critical value, the exchange spin interaction plays a dominant role, and the so-called \textit{compressible superfluid} ($CS$) state forms. From a study of the analytically tractable $XX$ models with a nearest-neighbors interaction \cite{Lieb1961}, it is well known that at $\Delta=0$, $CS$-state demonstrates zero minimum energy gap in the limit of infinite number of spins $N$. From Fig. \ref{fig:ED_gap_cut} one sees that the $CS$-state occurs also for the $XX$ model with a short/long-range exchange interaction and $x$-component of the magnetic field. The small wiggles observed in the dependence of $G(\tilde \Delta)$ are attributed to the finite-size effects (see the discussion below) \cite{Pasquale2008}. 

We also obtain the well-defined crossover between the $P$- and $CS$-states, and the critical value of $|\tilde \Delta_c(\beta)|$ separating these phases increases as the parameter $\beta$ decreases, which means that the range of the parameter $\Delta/J(1)$ for which the $CS$-state is observed, increases for a long-range interaction as $\beta \leq 2$. Our observations are summarized in the color plot of Fig. \ref{fig:ED_gap_2d}, where the minimum energy gap $G$ as a function of $\tilde \Delta$ and $\beta$ is presented. 
Notice here, that the effective spin model (\ref{eq:EffSpinHam-Long2}) with the nearest-neighbor interactions has been studied using the semi-classical product of states (no entanglement) approximation in \cite{kurmann1982antiferromagnetic}, and the lower bound of the critical value $|\tilde \Delta_c| > |\tilde \Delta_n|=2\sqrt{2}$ for the quantum phase transition between the $P$- and $CS$-states, was obtained. Using the Jordan-Wigner transformation and following mean-field analysis the quantum corrections resulting in a slight increase of the critical value $|\tilde \Delta_c| \simeq 3$ have been obtained in Refs. \cite{Ovchinnikov2002,Ovchinnikov_2_2002}. 
As $\beta \gg 1$, i.e., for short-range spin-spin exchange interaction, our numerical results obtained for the minimum energy gap $G$ dependence on the $\tilde \Delta$ in both $P$ and $CS$ collective quantum states are consistent with such analysis \cite{Ovchinnikov2002,Ovchinnikov_2_2002}. 

\begin{figure}
    \subfloat[]{
    \includegraphics[width=0.95\columnwidth]{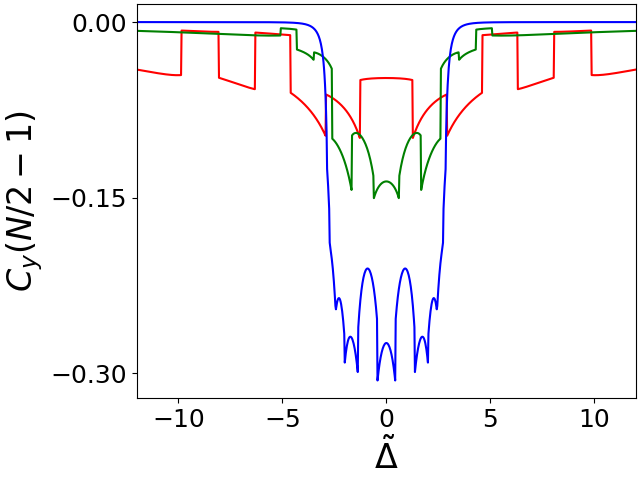}
    \label{fig:ED_y_correl_cut}}
    \\
    \vspace{-0.8\baselineskip}
    \subfloat[]{
    \includegraphics[width=0.95\columnwidth]{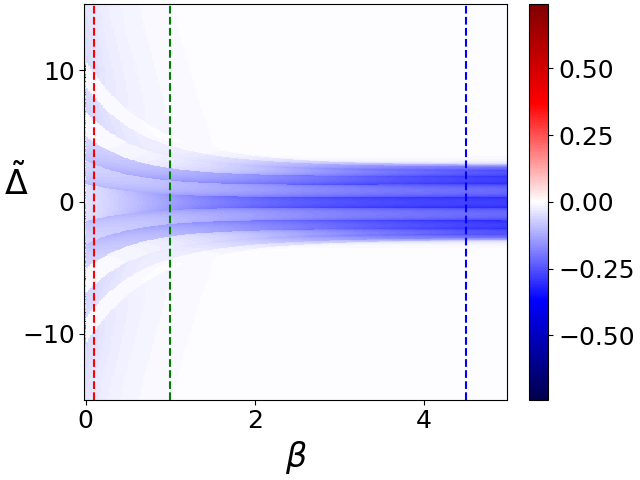}
    \label{fig:ED_y_correl_2d}}
    \caption{(color online) (a) Calculated dependence of the spatial correlations along $y$-axis characterized by $C_y(N/2-1)$ on the dimensionless parameter $\tilde \Delta$ for various values of $\beta$: $\beta = 4.5$ (blue line), $\beta = 1.0$ (green line), $\beta = 0.1$ (red line); 
    (b) The color plot of the spatial correlations $C_y(N/2-1)$ as the function of $\tilde \Delta$ and $\beta$. The color dashed lines correspond to the values of $\beta$ presented in Fig. \ref{fig:ED_y_correl_cut}. 
    The parameters were chosen as $N=12, \alpha = -0.8, E_J/E_C = 80$.}
\end{figure}

The \textit{spatial} properties of the collective states can be further characterized through the spatial correlation function of the $y$-component of local spins:
\begin{equation} \label{eq:SpatialCorrfunction}
C_y(|i-j|)=<\sigma_{y,i}\sigma_{y,j}>.
\end{equation}
The obtained dependencies of $C_y(N/2-1)$ on the parameter $\tilde \Delta$ for different values of $\beta$ are presented in Fig. \ref{fig:ED_y_correl_cut}. In the $P$-state, spatial correlations between spins located at a large distance are absent, and in the $CS$-state the negative spatial correlations oscillating with $\tilde \Delta$ were found. However, as the parameter $\beta \leq 2$ (a long-range exchange spin interaction), the spatial correlations in $y$-direction are strongly diminished. Thus, the collective phase obtained in the presence of a long-range exchange spin interaction can be called a \textit{weakly compressible superfluid} ($w$-$CS$) state. The complete dependence of $C_y(N/2-1)$ on the parameters $\tilde \Delta$ and $\beta$, is presented in Fig. \ref{fig:ED_y_correl_2d}. 

\begin{figure}
    \centering
    \includegraphics[width=0.95\columnwidth]{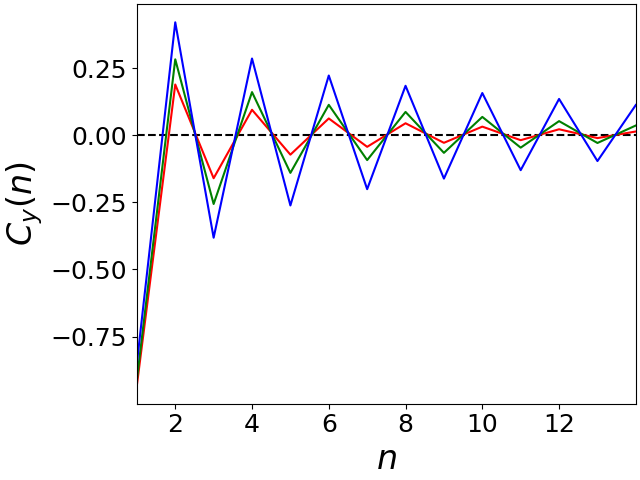}
    \caption{(color online) Calculated spatial correlation function $C_{y}(n)$ in the absence of the local quantum tunneling ($\tilde \Delta=0$) for different values of $\beta$: $\beta = 4.5$ (blue line), $\beta = 1.0$ (green line), $\beta = 0.1$ (red line). The other parameters were chosen as: $\alpha=-0.8, N=15$.}
    \label{fig:cor_y_d}
\end{figure}

\begin{figure}
    \centering
    \includegraphics[width=0.95\columnwidth]{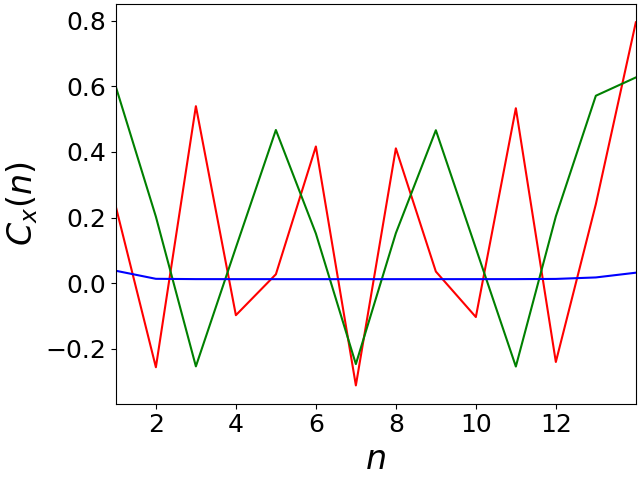}
    \caption{(color online) Calculated spatial correlation function $C_{x}(n)$ for non-zero local quantum tunneling $\tilde \Delta=4$ for different values of $\beta$: $\beta = 4.5$ (blue line), $\beta = 1.0$ (green line), $\beta = 0.1$ (red line). The other parameters were chosen as: $\alpha=-0.8, N=15$.}
    \label{fig:cor_x_d}
\end{figure}

The $CS$-state realized in a short-range interaction regime, as $\beta \gg 1$, also shows pronounced antiferromagnetic oscillations in the $C_y(n)$ dependence (see, the blue line in Fig. \ref{fig:cor_y_d}). In the $w$-$CS$-state, obtained in a long-range interaction regime ($\beta \leq 2$), the amplitude of the antiferromagnetic oscillations of $C_y(n)$ decreases (see, the green and red lines in Fig. \ref{fig:cor_y_d}). 
For a non-zero $\tilde \Delta$, the oscillatory behaviour of the correlation function with a dependence of the period and amplitude on $\tilde \Delta$ (similar to the observations in \cite{Ovchinnikov_2_2002}) and $\beta$ (see Fig. \ref{fig:cor_x_d}) persists. We also calculated the correlation function in $x$-direction 
\begin{equation}
    C_x(|i-j|) = <\sigma_{x,i} \sigma_{x,j}> - M_x^2, 
\end{equation}
where $C_x(n)$ becomes zero for large $\tilde \Delta$ in \textit{P}-state. In Fig. $\ref{fig:cor_x_d}$ we present $C_x(n)$ for $\tilde \Delta = 4$. $C_x(n)$ shows incommensurate behaviour, indicated previously in Ref. \cite{Ovchinnikov_2_2002}, and oscillations with large amplitude.

\begin{figure}
    \subfloat[]{
    \includegraphics[width=0.95\columnwidth]{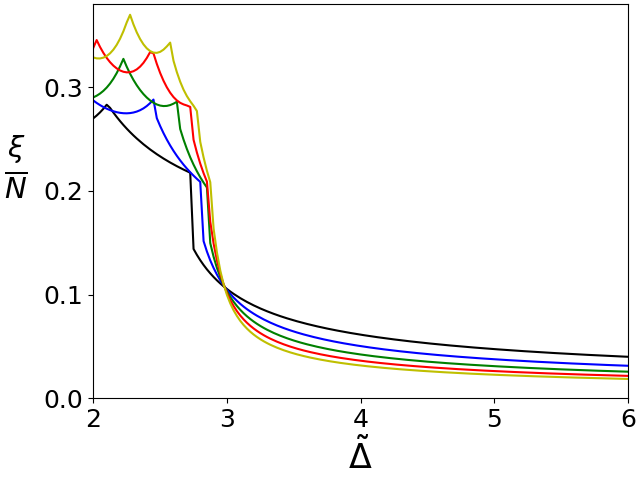}
    \label{fig:fss_bet4.5}}
    \\
    \vspace{-0.8\baselineskip}
    \subfloat[]{
    \includegraphics[width=0.95\columnwidth]{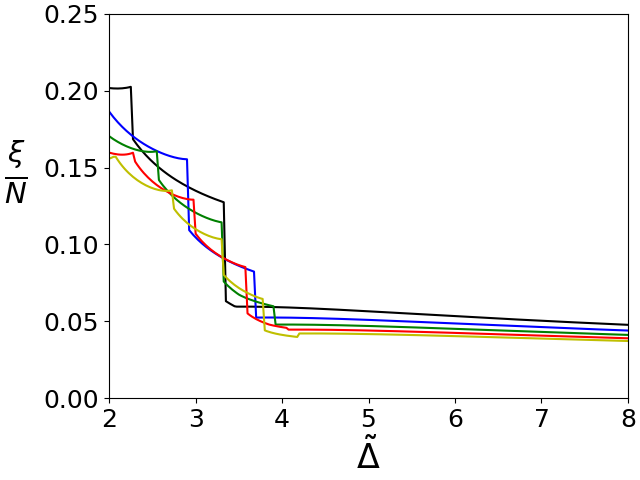}
    \label{fig:fss_bet1.5}}
    \caption{(color online) Claculated dependencies of the relative correlation length, $\xi/N$, on $\tilde \Delta$ for different values of the spin chain length, $N$: $N=6$ (black lines), $N=8$ (blue lines), $N=10$ (green lines), $N=12$ (red lines), $N=14$ (yellow lines). For a short-range interaction (a) ($\beta=4.5$) the typical scaling behavior as a single crossing point of different graphs, $\tilde \Delta \approx 3$ determines the phase transition between $P$- and $CS$ states. For a long-range interaction ($\beta = 1.5$) shown in (b) the scaling behavior is absent.    
    }
\end{figure}

To conclude this Section, we notice that in the $CS$ ($w$-$CS$)-states, the amplitude of oscillations of $C_y(n)$ exponentially decays with the distance $n$, and one can introduce the correlation length, $\xi$. Explicitly, we define the correlation length as (see Refs.~\cite{Sandvik2010})
\begin{equation}
    \xi = \frac{1}{q_1}\sqrt{\frac{S(0)}{S(q_1)} - 1},
    \label{correlation-length}
\end{equation}
where $q_1 = 2\pi/N$ and
\begin{equation}
    S(q) = \sum_{j=0}^N |C(j)|\cos(qj).
    \label{S-definition}
\end{equation}
The correlation length $\xi$ depends on the parameters $\tilde \Delta$, $\beta$, as well as the total number of spins $N$. In the $CS$ state, we obtain that the dependencies of the ratio $\xi/N$ on $\tilde \Delta$ for various $N$ demonstrate a standard scaling behavior \cite{Sandvik2010, Pang2019, asada2002anderson,Starkov2023}. It is presented in Fig. \ref{fig:fss_bet4.5} for $\beta=4.5$, where one can see all curves intersecting in a single point. Thus, these scaling arguments support the conclusion that $\tilde \Delta \approx 3$ determines the phase transition between $P$ and $CS$ states \cite{Ovchinnikov2002,Ovchinnikov_2_2002,kurmann1982antiferromagnetic}.

On the contrary, in the $w$-$CS$ the dependencies of $\xi/N$ on $ \tilde \Delta$ do not show a single intersection point (see, Fig. \ref{fig:fss_bet1.5}), and therefore, to obtain the critical value $\tilde \Delta$ determining the phase transition between $w$-$CS$ and $P$ states, one has to go beyond the direct diagonalization procedure.

\section{\label{sec6:level1}Conclusions}
In conclusion, we theoretically study the collective quantum phases occurring in frustrated saw-tooth chains of Josephson junctions embedded in a dissipationless transmission line. The frustration was introduced through the periodic arrangement of $0$- and $\pi$-Josephson junctions. 
The frustrated regime is realized for the parameter $-1<\alpha<-0.5$,  where $\alpha E_J$ is the Josephson coupling energy of $\pi$-Josephson junctions. In this regime a single basic cell of the system, i.e., a superconducting triangle, composed of two $0$- and one $\pi$-Josephson junctions embedded in the superconducting loop, shows two stable states separated by a potential barrier. These two stable states correspond to the persistent currents (magnetic vortex/antivortex) flowing in clockwise or anticlockwise directions. In the quantum frustrated regime, the macroscopic quantum tunnelling yields coherent quantum beats between these two states. Thus, a single basic element of a frustrated saw-tooth chain of Josephson junctions is equivalent to a single flux qubit biased at the symmetry point \cite{Orlando1999}. 

The collective macroscopic quantum dynamics of a \textit{large} frustrated saw-tooth chain of Josephson junctions crucially depends on the interaction between Josephson junctions located in different cells. Direct embedding of $\pi$-Josephson junctions in the transmission line establishes either a long or short-range interaction. The range of interaction is determined as $\ell_0 \simeq \sqrt{C/C_0}$ where $C$ and $C_0$ are capacitances of Josephson junctions and the transmission line, accordingly. 

By making use of the variational approach, we reduce the quantum superconducting circuit Hamiltonian of a system to the effective interacting spin chain $XX$ Hamiltonian (Eq. (\ref{eq:EffSpinHam-Long})), in which the local spin-flips and a long/short exchange spin-spin interaction are taken into account. We obtain that the amplitude of a local spin-flips $\Delta$ is drastically suppressed in the limit of long arrays, $N \gg \ell_0$, and therefore, the dimensionless parameter $\tilde \Delta=\Delta/J(1)$ determining the appearance of a specific collective quantum phase can be tuned in a wide region by changing the physical parameters $E_J$, $\alpha$, $C$, $C_0$ and $N$. The exchange spin-spin interaction decays with distance as a power law with exponent $\beta$, which can be either larger or smaller than $1$. 

Using the direct numerical diagonalization of the Hamiltonian (Eq. (\ref{eq:EffSpinHam-Long2})) for the spin chains up to $N=16$ sites, we identify several collective quantum phases: the paramagnetic ($P$)-state, the compressible superfluid ($CS$) and weakly compressible superfluid ($w-CS$) states, characterized by absence (presence) of the minimum energy gap (see, Fig. \ref{fig:ED_gap_cut}), and substantial spatial antiferromagnetic correlations (see, Fig. \ref{fig:ED_y_correl_cut}). Varying the parameters $\tilde \Delta$ and $\beta$, we explored the complete phase diagram (see, Figs. \ref{fig:ED_gap_2d} and \ref{fig:ED_y_correl_2d}). 

Finally, we notice that an application of a small magnetic field allows us to include local $\hat \sigma_{z,n}$-terms in the spin interacting Hamiltonian (\ref{eq:EffSpinHam-Long}). We anticipate that the quasi-1D $f$-JJAs directly embedded in the low dissipative transmission line can be used as a convenient experimental platform to establish the quantum simulations of large strongly interacting spin systems, and to obtain various quantum collective states.

\begin{acknowledgments}
We acknowledge the support provided by the Konrad-Adenauer-Stiftung, Germany.
\end{acknowledgments}

\appendix

\section{Higher-order multiple simultaneous spin flips}
Here, we verify the statement that the amplitudes of higher-order spin flips are exponentially small and can be neglected in the effective Hamiltonian (\ref{eq:EffSpinHam-Long}). The amplitude of the process involving spin flips in $n$ cells depends on the distances between the constituent spins and the \textit{parity } of each spin-flip pair. First, the amplitude of an \textit{even} two cells ($n=2$) spin flips
\begin{equation}
    J^+(|n-m|) = J_0 \exp \lbrace -4 u_0^2 \left[ A(0) + A(|n-m|) \right]  \rbrace .
\end{equation}
is exponentially smaller than the odd amplitude $J(n-m)$ (see, Eq. ($\ref{Twospinsflip-longJJA}$)): 
\begin{equation}
    \frac{J^+(|n-m|)}{J(|n-m|)} = \exp(-8 u_0^2 A(|n-m|)).
\end{equation}
The higher-order terms for $n\geq 2$ can be written as 
\begin{align}\label{higher-order-amplitudes}
& \frac{J^{(n)}(\lbrace \mathcal{P}_i, d_i \rbrace)}{J_0^{(n)}} =  \nonumber \\ 
& \exp \lbrace -4 u_0^2 \left[ \frac{n A(0)}{2}  + \sum_{i=1}^{\binom{n}{2}} \mathcal{P}_i A(d_i) \right] \rbrace,
\end{align}
where $\mathcal{P}_i = \lbrace+1, -1 \rbrace$  is the parity and $d_i$ the distances between the two constituents of the $i$-th spin flip. All the $ \lbrace \mathcal{P}_i, d_i \rbrace $ are fixed for a given spin flip configuration. The sum $i$ runs over all possible spin flips (spin pairs). 

As $A(d_i)$ decays strongly with increasing distance $d_i$, we consider only clusters of spin flips ($\lbrace d_i = 1,2 \rbrace$). Thus we get the third order term 
\begin{align} \label{eq:highorderterm}
    & \frac{J^{(3)}}{J^{(3)}_0} = \nonumber \\
    & \exp\lbrace -4u_0^2 \left[ \frac{3 A(0)}{2} \pm  A(1) \pm A(1) \pm  A(2) \right] \rbrace. 
\end{align}
Such expression can be easily extended to higher orders. Since the exponent in (\ref{eq:highorderterm}) increases logarithmically with $N$ the all terms $J^{(n)}$ with odd $n$ are strongly suppressed.

\begin{figure}
    \centering
    \includegraphics[width=.49\textwidth]{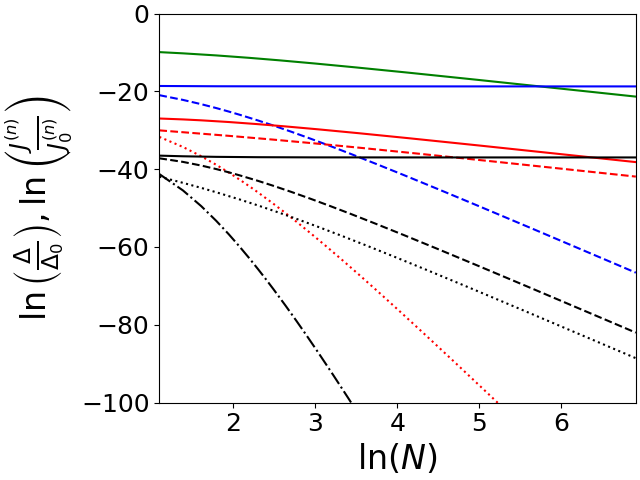}
    \caption{(color online) Comparison of the terms up to $4$th order: 1st order (green), $2$nd order (blue), $3$rd order (red), $4$th order (black). The amplitude of quantum tunneling $\Delta$ (green solid line), the odd second order exchange interaction strength $J(1)$ (blue solid line), the even order term $J^+(1)$ (blue dashed line). The third order combinations are given as follows: odd-odd (red solid line), odd-even (red dashed line) and even-even (red dotted line). For the 4th order (black), we have the following combinations: o-o-o (solid), o-o-e (dashed), o-e-e (dotted) and e-e-e (dashdot).}
    \label{fig:high_order_sf}
\end{figure}

The fact that the higher-order terms with $n>2$ are indeed exponentially small can also be seen in Fig. \ref{fig:high_order_sf}, which shows all spin flip amplitudes up to $4$th order for a connected cluster of spin flips. From this, we conclude that the highest amplitude within a certain order occurs when all short ranged (nearest-neighbor) spin flips have odd parity. Furthermore, if the order is increased, we find that the amplitude is exponentially small and can be neglected in the effective model. 

\bibliography{article}

\end{document}